\documentclass[prl,aps,floats,letterpaper,floatfix,nofootinbib,preprintnumbers,twocolumn]{revtex4}
\usepackage{amsmath}
\usepackage{color}
\usepackage[mathenv]{}
\usepackage{graphicx}
\usepackage{natbib}
\begin{document}
\bibliographystyle{apsrev}

%\title{Observation of parametric instability and opto-mechanical rigidity
%in a suspended Fabry-Perot cavity}

\title{Measurement of radiation-pressure-induced optomechanical
dynamics in a suspended Fabry-Perot cavity}

\author{Thomas Corbitt}
\author{David Ottaway}
\author{Edith Innerhofer}
\author{Jason Pelc}
\author{Nergis Mavalvala}

\affiliation{LIGO Laboratory, Massachusetts Institute of Technology, Cambridge, MA 02139, USA}

\begin{abstract}
We report on experimental observation of radiation-pressure
induced effects in a high-power optical cavity. These effects play
an important role in next generation gravitational wave (GW)
detectors, as well as in quantum non-demolition (QND)
interferometers. We measure the properties of an optical spring,
created by coupling of an intense laser field to the pendulum mode
of a suspended mirror; and also the parametric instability (PI)
that arises from the nonlinear coupling between acoustic modes of
the cavity mirrors and the cavity optical mode. Specifically, we
measure an optical rigidity of $K = 3 \times 10^4$~N/m, and PI
value $R = 3$.
\end{abstract}

%\pacs{04.80.Nn, 03.65.ta, 42.50.Dv, 95.55.Ym}
\definecolor{purple}{rgb}{0.6,0,1}
\preprint{\large \color{purple}{LIGO-P050045-00-R}}
\date{\today}
\maketitle

% \section{Introduction}
Second generation interferometric gravitational wave (GW)
interferometers are anticipated to have in excess of 0.5 MW of
circulating power in the long Fabry-Perot arm cavities
\cite{pfspie}. High circulating power is needed to reduce the
effect of photon counting statistics (shot noise) which limits the
performance of these instruments at higher frequencies (above
$~200$ Hz). In this high power regime, the assumption that the
dynamics of the suspended masses and the light field used to
measure their motion can be treated separately no longer applies,
and a rich variety of physical phenomena result from radiation
pressure effects. The radiation pressure effects include
parametric instability (PI)
~\cite{MSUpla2001,CITpla2002,MSUpla2002,zhao}, optical tilt
instability~\cite{sidles_sigg}, quantum radiation pressure noise
and the opto-mechanical rigidity~\cite{BC1,BC3,code_paper}. In
this Letter we report on the observation and characterization of
two radiation pressure induced phenomena in a detuned Fabry-Perot
cavity with mirrors suspended as pendulums: the optical spring
effect and a PI. These measurements are in a frequency, mass and
optical rigidity regime that differ from previous measurements by
many orders of magnitude, and are of particular importance to GW
detectors and quantum non-demolition (QND) devices.

The optical spring effect occurs in Fabry-Perot cavities that are
detuned from resonance. When a cavity with movable mirrors is
detuned, radiation pressure creates an optical restoring force
that can significantly increase the force required to change the
separation between two suspended cavity mirrors. This increased
rigidity, and the associated optical spring, modifies the response
function of the mirror modes. At frequencies below the optical
spring resonant frequency, the response of cavity length to
external disturbances (e.g., driven by seismic and/or thermal
forces) is suppressed by the increased rigidity factor. This
suppression factor makes the optical spring an important feature
of QND interferometers~\cite{code_paper,Corbitt}. The optical
spring effect also occurs in Advanced LIGO~\cite{BC1}. The
response of these opto-mechanically stiffened oscillators is also
predicted to be unstable~\cite{BC3}.

The optical spring effect has been previously demonstrated on a
$1.2$ g mass held in a flexure mount~\cite{Sheard}. In that
experiment the mechanical resonance of the mass/flexure structure
at $303$ Hz was altered by $3\%$ with the application of radiation
pressure force, corresponding to an optical rigidity of about $150
{\rm ~N}/{\rm m}$. In the experiment reported here, $0.25$ kg
mirrors are suspended as pendulums with a resonant frequency of
$1$ Hz for the longitudinal mode (motion along the optic axis of
the cavity). With detuning of the cavity, the resonance is shifted
upwards by nearly $2$ orders of magnitude. We also confirm the
unstable nature of the resonance.

% \subsection{Parametric instability}
In recent years there has been some debate on the potential of PIs
to adversely impact the performance of second generation GW
interferometers. The risk of PIs in future detectors arises from
the high circulating power and the low mechanical loss (high
quality factor, or $Q$) materials planned for use in future test
masses. High mechanical $Q$ materials are required to limit the
effect of thermal noise on the sensitivity of the
device~\cite{saulson}.

The PI process is as follows: light circulating inside a
Fabry-Perot cavity is phase modulated due to the vibrations of a
mirror mechanical mode. The phase modulation creates a pair of
sidebands equally spaced on either side of the carrier at the
mechanical mode frequency. If the cavity has an asymmetric optical
response to these sidebands, one will experience a greater
build-up than the other, and they will not be balanced. An
imbalance in these sidebands will result in a fluctuating
amplitude and hence an oscillating radiation pressure force. If
the upper sideband (anti-Stokes mode) is favored then cold damping
of the mechanical mode occurs, in which the mechanical quality
factor of the mode is reduced. If the lower sideband (Stokes mode)
is favored, a run-away process may occur when the rate at which
the mechanical mode is pumped by the Stokes mode is greater than
the rate that energy is lost by mechanical dissipation.

Braginsky et al.~\cite{MSUpla2001} first reported on the danger of
PIs in high power GW detectors. They suggested that for
kilometer-scale cavities with high circulating power and low free
spectral range (fsr), a Stokes mode at the fsr could cause the
cavity to become unstable. They also warned that the density of
mechanical modes around the fsr of the cavity could overlap with a
higher order mode of the cavity, leading to instability.
Application to realistic interferometers by Zhao et al.
\cite{zhao} has confirmed that there are likely to be modes with
sufficient parametric gain to be unstable.

Parametric instabilities have been observed in resonant bar
detectors with microwave resonator readouts \cite{Cuthbertson} and
in optical micro-cavities \cite{Kippenberg}. Kippenberg et al.
observed radiation-pressure-induced parametric oscillation
instabilities in ultrahigh-Q toroidal optical microcavities, at
frequencies of 4.4 to 49.8 MHz and modal masses of $10^{-8}$ to
$10^{-9}$ kg. The experiment reported in this Letter differs from
these experiments in that it demonstrates PI at $28.188$ kHz, in a
suspended cavity apparatus with an effective mass of $0.23$ kg.
The mass and frequency regime of this experiment are of particular
relevance to GW detectors~\cite{pfspie} and ponderomotive
squeezing experiments \cite{code_paper,Corbitt}.

% \section{Unified theory for optical spring and parametric instability}
Since the PI and optical spring described in
this Letter both arise from radiation pressure of the fundamental
mode of the intracavity field, we present a simple unified model
for both effects. Our model gives similar results to that of
Braginsky et al.~\cite{MSUpla2002}, who define a dimensionless
PI susceptibility, $R$. Since the level of
PI depends on the relative excitation of the
Stokes and anti-Stokes mode, Braginsky et al. consider both the
cold damping effect of the anti-Stokes and parametric
amplification of the Stokes mode.

In our experiment, the Stokes and anti-Stokes modes experience
different optical gains. They occur under the same optical
resonance, but one that is detuned for the carrier frequency. The
PI occurs because a Stokes mode in the TEM$_{00}$ spatial mode
beats with the carrier in the same spatial mode, and interacts
with the acoustic mode of the mirror. Because our experiment does
not involve higher order spatial modes, we can construct a simple,
unified model that describes both the optical rigidity and the PI.
Defining $s = j\,\Omega$, where $\Omega$ is the measurement
frequency, the optical rigidity is given by
~\cite{code_paper,Corbitt}

\begin{equation}
K\left(s\right) =
-K_0\,\frac{\gamma^{2}}{\left(\gamma+s\right)^2+\delta^2},
\end{equation} where
\begin{eqnarray}
\label{eq:Theta} K_0 = \frac{4\omega_0\, I_0\, \left(\delta/\gamma
\right) }{
c^2}\left[\frac{4}{\mathcal{T}_I}\,\frac{1}{1+\left(\delta/\gamma\right)^2}
\right]^2 \,,
\end{eqnarray}

\noindent and $\gamma$ is the cavity linewidth, $\delta$ is the
detuning of the cavity from resonance, $I_0$ is the power incident
on the cavity, and ${\cal T}_I$ is the transmission of the cavity
input mirror. The response of the mirror to a driving force near a
resonance may be modelled as a simple harmonic oscillator with a
mechanical resonant frequency $\Omega_0$,
\begin{equation}
P\left(s\right) = \frac{1}{M_{{\rm eff}}} \times
\frac{1}{s^2+\Omega_0^2 + s \, \left(\Omega_0/Q_m\right)}.
\label{eq:P}
\end{equation} We define the effective mass according to
\begin{equation}
M_{{\rm eff}} \equiv  \frac{Q_m}{\Omega_0^2 {\cal R}},
\end{equation}
where $\cal R$ is the displacement per force at $\Omega_0$,
averaged over the laser beam. The motion of the mirror surface
creates a phase shift of the light in the cavity. Since the cavity
is detuned from resonance, the unbalanced propagation of the upper
and lower sidebands is converted into an intensity modulation of
the intracavity power. This, in turn, pushes back against the
mirror surface. The optical rigidity, therefore, forms an optical
feedback system, with a modified response
\begin{equation}
P\left(s\right)^\prime = \frac{P\left(s\right)}{1-P\left(s\right)K\left(s\right)}.
\end{equation}
Though it is not strictly valid in our experiment, it is
illustrative to consider the case where $s\ll \gamma$ and $\delta
\ll \gamma$:
\begin{equation}
P\left(s\right)^\prime \approx \frac{1}{M_{{\rm eff}}} \times
\frac{1}{s^2+\Omega_0^{\prime 2} + s \,\left(
\Omega_0^\prime/Q^\prime\right)}, \label{eq:Pprime} \end{equation}
where
\begin{equation}
\Omega_0^{\prime\, 2} = \Omega_0^2 + \frac{K_0}{M_{{\rm
eff}}}\,,\end{equation}
%\begin{equation} \frac{Q^\prime}{Q_m} =
%\left[ \left(1+\frac{K_0}{M_{{\rm eff}}\Omega_0^2}\right)^{-1/2} -
%\frac{2\,K_0\,Q_m}{M_{{\rm
%eff}}\gamma\sqrt{\Omega_0^2+\frac{K_0}{M_{{\rm
%eff}}}}}\right]^{-1}\,.
%\end{equation}
\begin{equation} \frac{Q^\prime}{Q_m} =
\left(1+\frac{K_0}{M_{{\rm eff}}\Omega_0^2}\right)^{1/2} \left[ 1
- \frac{2\,K_0\,Q_m}{M_{{\rm eff}}\gamma\Omega_0}\right]^{-1}\,.
\end{equation}
From Eqs.~(\ref{eq:P}) and (\ref{eq:Pprime}), we see that the
response of the closed system is identical to that of a harmonic
oscillator, with a modified resonant frequency $\Omega_0^{\prime}$
and quality factor $Q^{\prime}$. The $R$ value defining the
stability of the system, corresponding to definition in
Ref.~\cite{MSUpla2002}, can be found by the relations
\begin{equation}Q^\prime = \frac{Q_m}{1-R} \hspace{0.5cm} {\rm or}
\hspace{0.5cm} \tau^\prime = \frac{\tau_m}{1-R}\,.\label{eq:tau}
\end{equation} Here $\tau^\prime$ and $\tau_m$ refer to
the opto-mechanical and mechanical timescale for the ringing of
the mode, respectively. In our system, we consider two modes of
the mirror in two different frequency regimes. The first mode is
the pendulum motion of the mirrors. For this case, the optical
rigidity is much greater than the gravitational restoring force,
such that $K_0 \gg M_{{\rm eff}}\Omega_0^2$, $\,\Omega_0^\prime
\approx \sqrt{K_0/M_{{\rm eff}}}$, and $Q^\prime \approx
-\frac{\gamma}{2\sqrt{K_0 / M_{{\rm eff}}}}$, which gives
\begin{equation}
R \approx 1+\frac{2 Q_m\sqrt{K_0 / M_{eff}}}{\gamma}\,.
\end{equation}
We note that the resonant frequency is shifted ($\Omega_0^\prime
\rightarrow \Theta$), and the resonance is inherently unstable ($R
> 1$), with a quality factor $Q^{\prime}$ independent of $Q_m$.

The other mode is the acoustic drumhead motion of the mirror. For
this mode, the optical restoring force is much less than the
mechanical restoring force, such that $K_0 \ll M_{{\rm
eff}}\,\Omega_0^2$ and $\Omega_0^\prime \approx \Omega_0$, which
gives
\begin{equation}
R \approx \frac{2\,K_0\,Q_m}{M_{{\rm eff}}\gamma\,\Omega_0}.
\label{eq:R}
\end{equation}
For this mode, the resonant frequency does not change, but the $Q$
of the mode may be altered, and may even be made negative so that
the mode becomes unstable.

%\section{Experiment and results}

\begin{figure}[t]
\begin{center}
\includegraphics[width=8cm]{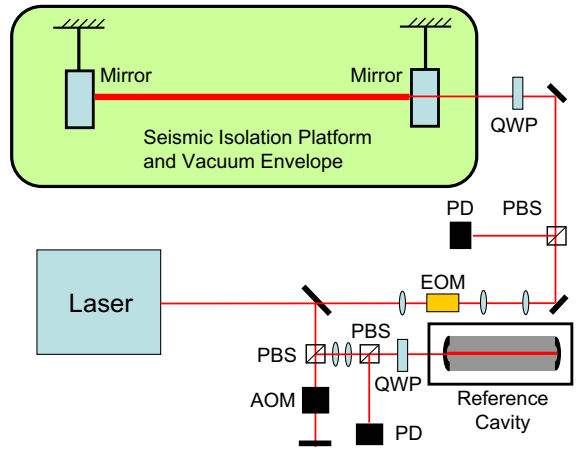}
\caption{Schematic representation of the experiment, showing the 1
m long Fabry-Perot cavity suspended in vacuum. Abbreviations are
acoustooptic modulator (AOM), electrooptic modulator (EOM),
photodiode (PD), polarizing beam splitter (PBS), and quarter-wave
plate (QWP). Most of the 10 W output of the 1064 nm Nd:YAG laser
light is directed to the suspended cavity. A small fraction of is
pick-off and frequency shifted by double passing through the AOM.
The frequency shifted light is used to lock the laser frequency to
the reference cavity resonance with high bandwidth ($>500$ kHz).
Frequency shifting via the AOM provides a means of high speed
frequency modulation of the stabilized laser output; it is an
actuator for further stabilizing the laser frequency with
reference to the suspended cavity. The input mirror of the cavity
has a transmission of $0.63\%$, giving a linewidth of $75$ kHz.
Not shown are feedback loops actuating on the cavity length and
laser frequency (via the AOM). \label{fig:schematic}}
\end{center}
\end{figure}

The experimental set-up used to observe the PI and the
opto-mechanical rigidity is shown in Fig.~\ref{fig:schematic}. The
experiment consists of a Fabry-Perot cavity comprising two 0.25 kg
mirrors suspended by single loops of wire. The suspended mirrors
are located in a vacuum chamber and are mounted on a three-layer
passive vibration isolation system. The motion of the mirrors is
controlled by forces due to small current-carrying coils placed
near magnets that are glued to the back surface of each mirror.

Light from a frequency-stabilized laser is resonated inside the
suspended Fabry-Perot cavity. Approximately $3.6 $ Watts of laser
power is injected into the cavity. The Pound-Drever-Hall locking
technique is used to hold the cavity on resonance (or detuned,
with an injected offset).

\begin{figure}[t]
%\begin{center}
%\begin{tabular}{cc}
\includegraphics[width=0.45\textwidth]{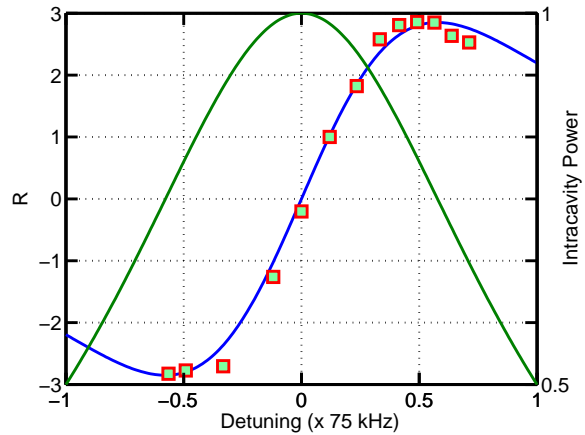}
%\includegraphics[width=0.45\textwidth]{R_vs_power}
%\end{tabular}
%\end{center}
\caption{The instability measure $R$ is plotted as a function of
detuning of the cavity from resonance, for fixed input power. The
solid (blue) curve is the theoretical prediction, according to
Eq.~\ref{eq:R}, with fit parameter $M_{{\rm eff}}$; the measured
data are shown as colored squares. Also shown is the power built
up in the cavity, normalized to unity (green curve).
\label{fig:R}}
\end{figure}

To study the PI, we couple the drumhead mode at $28.188$ kHz of
one mirror to the optical fields. To measure the value of $R$ for
this mode as a function of detuning and power, it is necessary to
measure ringup and ringdown times ($\tau^\prime$ and $\tau_m$ in
Eq.~\ref{eq:tau}) of the mode for various detunings and input
power levels. To ensure that the control system that keeps the
cavity on resonance (or at the detuning point) does not interfere
with the measurements, the length of the cavity is locked to the
frequency of the laser with approximately $1$ kHz bandwidth. A
$-60$ dB notch filter at $28.188$ kHz is added to the servo to
eliminate any interference of the servo system with the
measurement of the drumhead motion of the mirror. For the cases in
which $R>1$, we allow the mode to begin oscillating and capture
the ringup of the mode and fit an exponential to find the
timescale, and therefore the quality factor for the mode, which in
turn gives the value of $R$. To measure values at $R<1$, we first
detune the cavity to a point where $R>1$ and allow the mode to
ringup, then quickly detune to the desired point and capture the
ring down, and fit an exponential decay to find the value of $R$.
The results of the measurements of $R$ at various detunings, for a
fixed input power, are shown in Fig.~\ref{fig:R}. The value of
$M_{{\rm eff}}$ was treated as a free parameter that was fit to
the data; we found a value of $0.227$~kg. The measurements show
good agreement with the predicted values. We also measured $R$ as
a function of cavity power, with fixed detuning (set to 75\% of
the maximum power), and established the linear dependence on
power, also showing that $R$ goes to zero at zero injected power.

%The value for $M_{{\rm eff}}$ may also be calculated using finite
%element analysis, and gives a value of [need value here, and cite
%Dennis].

\begin{figure}[tbh]
\includegraphics[width=9cm]{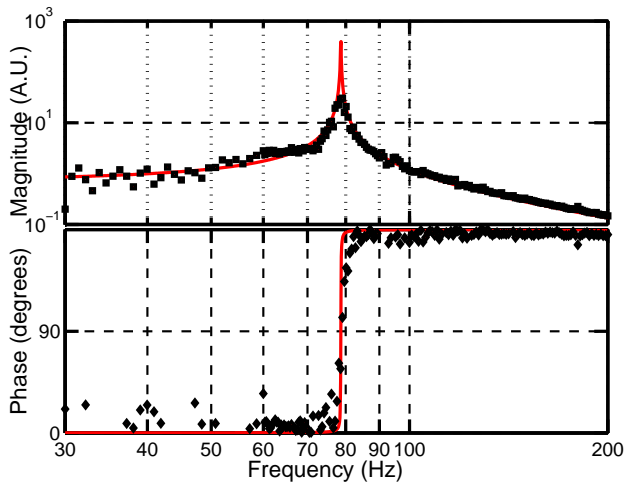}
\caption{The transfer function of applied force to displacement of
the end mirror of the cavity, showing the optical spring
resonance. The solid (red) curve is the theoretical prediction
with no free parameters, based on measurement of the intracavity
power, and the quadrangles are the measured data. The phase
increases from 0 to 180 degrees at the resonance, showing that the
resonance is unstable. \label{fig:OS}}
\end{figure}

Measuring the optical spring presented some challenges. The
instability of the drumhead mode must be stabilized before the
measurement can be performed. To accomplish this stabilization, we
modify our servo so that we tune the frequency of the laser to
follow the length of the cavity at frequencies from $~300$ Hz to
$50$ kHz. The modified servo system suppresses the formation of
the $28.188$ kHz sidebands that are an integral part of creating
the PI. With sufficient suppression at $28.188$ kHz, the drumhead
mode will not be excited. The second difficulty is that the
intrinsic motion of the mirrors at low frequencies is quite large
and requires a strong servo system to hold the cavity on
resonance. This has the unfortunate consequence of placing the
optical spring resonance within the bandwidth of the servo system.
The gain of the servo system explicitly depends on
$P\left(s\right)^\prime$, however, so by characterizing the servo
system, $P\left(s\right)^\prime$ is also characterized and the
optical spring may be measured. The results from this measurement
are shown in Fig.~\ref{fig:OS}. We note that the transfer function
is similar to that of a simple harmonic oscillator, with one
important exception: that the resonance shows a negative damping
constant (the phase {\it increases} by $180^{\circ}$ at the
resonance), as predicted by our model. We suspect that the
smearing out of the sharp predicted peak in the data is caused by
fluctuations of the intracavity power. The measured response is,
however, consistent with the theoretical prediction, with no free
parameters. From the measured frequency of the optical spring
resonance, we infer the optical rigidity to be $K = 3 \times 10^4
{\rm ~N}/{\rm m}$.

% \section{Conclusions}
In summary, we have measured both the optical spring effect and a
PI in a high-power Fabry-Perot cavity with mirrors suspended as
pendulums. We find a maximum optical rigidity of $K = 3 \times
10^4 {\rm ~N}/{\rm m}$, and a PI with $R = 3$. Since the optical
rigidity is a crucial element of an experiment to generate
squeezed states using ponderomotive
rigidity~\cite{code_paper,Corbitt}, good agreement of our
experiment with theory is valuable. The PI is an unwanted effect
in that experiment, and its characterization and subsequent
damping are important for the next phase of experimentation with
higher finesse cavities and lighter mirrors. The PI identified and
damped in this Letter is an extension of the work of the Vahala
group~\cite{Kippenberg} to much lower frequencies (by 4 to 5
orders of magnitude) and much higher effective mass (by 9 to 10
orders of magnitude). Similarly, the optical rigidity created in
this experiment exceeds that of Sheard et al.~\cite{Sheard} by 2
orders of magnitude. The measurements in this Letter are an
important mass, frequency and optical rigidity regime for GW
detectors and QND interferometers.

We would like to thank our colleagues at the LIGO Laboratory,
especially D. Shoemaker for helpful comments on the manuscript. We
gratefully acknowledge support from National Science Foundation
grants PHY-0107417 and PHY-0457264.

%%%%%%%%%%%%%%%%%%%%%%%
%%%%%bibliography%%%%%%

\end{document}